\documentclass[smallextended]{svjour3} 
\smartqed
\usepackage{natbib}
\usepackage{mathptmx}
\usepackage{graphicx}
\usepackage{amssymb,amsfonts,amsmath}
\journalname{Journal of Molecular Evolution}
\begin{document}

\title{Membrane environment imposes unique selection pressures on transmembrane domains of G protein-coupled receptors}
\author{Stephanie J. Spielman \and Claus O. Wilke}
\titlerunning{Molecular evolution of structural domains in G protein-coupled receptors} 

\maketitle

\institute{S.J. Spielman and C.O. Wilke\at 
The University of Texas at Austin, Austin, TX 78731, USA.\\
Tel.: +1-239-877-1129\\
\email{stephanie.spielman@utexas.edu}\\
}

\bigskip
\noindent
$^*$Corresponding author\\
$\phantom{^*}$Email: stephanie.spielman@utexas.edu\\
$\phantom{^*}$Phone: 239-877-1129\\

\bigskip

\bigskip

\begin{abstract} 
We have investigated the influence of the plasma membrane environment on the molecular evolution of G protein-coupled receptors (GPCRs), the largest receptor family in Metazoa. In particular, we have analyzed the site-specific rate variation across the two primary structural partitions, transmembrane (TM) and extramembrane (EM), of these membrane proteins. We find that transmembrane domains evolve more slowly than do extramembrane domains, though TM domains display increased rate heterogeneity relative to their EM counterparts. Although the majority of residues across GPCRs experience strong to weak purifying selection, many GPCRs experience positive selection at both TM and EM residues, albeit with a slight bias towards the EM. Further, a subset of GPCRs, chemosensory receptors (including olfactory and taste receptors), exhibit increased rates of evolution relative to other GPCRs, an effect which is more pronounced in their TM spans. Although it has been previously suggested that the TM's low evolutionary rate is caused by their high percentage of buried residues, we show that their attenuated rate seems to stem from the strong biophysical constraints of the membrane itself, or by functional requirements. In spite of the strong evolutionary constraints acting on the transmembrane spans of GPCRs, positive selection and high levels of evolutionary rate variability are common. Thus, biophysical constraints should not be presumed to preclude a protein's ability to evolve.

\end{abstract}

\section{Introduction}
\label{intro}
A protein's evolution may be constrained by various functional or biophysical requirements. Membrane proteins, in particular, should be heavily constrained by the hydrophobic environment inside the membranes where they reside, specifically with regards to their transmembrane (TM) domains. This structural constraint biases amino acids found in TM domains towards non-polar, or hydrophobic, residues; polar amino acids comprise roughly 60\% of TM segments, compared to a 30\% frequency in extramembrane (EM) regions, whereas polar amino acids make up a mere 5\% of the TM \citep{Tourasse2000}. Although a protein's amino-acid composition is not a robust determinant of evolutionary rate, the underlying biophysical constraints yielding this bias presumably enforce a lower rate of evolution in TM regions relative to globular proteins or to EM regions of the same protein  \citep{Tourasse2000, Julenius2006}. The high concentration of buried residues in TM domains has additionally been proposed to be a dominant contributor to their low evolutionary rate \citep{Stevens2001,Oberai2009}, as highly buried protein residues are known to correlate with low evolutionary rates \citep{FranzosaXia2009,Ramseyetal2011}. 

Although the general patterns associated with membrane evolution have been loosely characterized, the evolutionary variability within the TM and EM spans, particularly across individual residues, is largely unknown. Previous studies investigating the evolution of membrane proteins have focused primarily on average evolutionary rates, neither addressing rate heterogeneity nor site-based evolutionary parameters \citep{Tourasse2000,Gilad2000,Clark2003,Julenius2006}. Additionally, those studies used either orthologous pairs or trios of sequences, which hindered statistical robustness and precluded any analysis of site-rate variation due to a dearth of data \citep{Tourasse2000,Gilad2000,Clark2003,Julenius2006}.

To obtain a more complete picture of membrane protein evolution, we have analyzed the evolutionary constraints acting on one of the most diverse membrane protein gene families in Metazoa, the G protein-coupled receptor (GPCR) family. GPCRs are the frequent targets of structural and biochemical studies; over 40\% of pharmaceuticals target GPCRs, and a multitude of diseases are caused by mutant GPCRs \citep{Dorsam2007,Schoneberg2004,Kristiansen2004,Fredriksson2003}. Phylogenetic analyses have shown that GPCRs form five main families, with the vast majority of human receptors belonging to the Rhodopsin-like (Family A) clade \citep{Fredriksson2003,Fredriksson2005}. Owing to their enormous diversity of biological functions and the ongoing expansion of their ligand repertoire, GPCRs have been described as one of the most evolutionarily successful gene families \citep{Bockaert1999,Lagerstrom2008}. Although protein sequences among, and indeed within, GPCR families are widely divergent, all GPCRs share a common structure characterized by a N-outside C-inside orientation with seven TM alpha helices spanning the plasma membrane, separated by three intracellular and three extracellular loops. 

GPCRs accept a wide variety of ligands, both endogenous (e.g.\ hormones, amines, or ions) and exogenous (e.g.\ odorants), and facilitate signal transduction through a G protein mediated pathway \citep{Kristiansen2004,Lagerstrom2008,Rosenbaum2009}.
Although some larger ligands do bind the extracellular portion of GPCRs, nearly all family A GPCRs, as well as many members of other GPCR families, bind ligands within their TM \citep{Vaidehi2002,Kristiansen2004,Bywater2005,Surgand2006,May2006,Park2008}. The notable expections to this trend are family C GPCRs, whose ligand-binding domains lie primarily in their extensive and diverse N-termini \citep{May2006,Park2008,Lagerstrom2008}. However, allosteric  modulators acting on all GPCR families bind within the TM. This commonality highlights the key role that the TM plays in the regulation of protein activity \citep{May2006,Lagerstrom2008}.

The TM domain is also a critical determinant of a GPCR's conformational state. Mutational studies have shown that altering specific residues in GPCR TM spans results in structural modifications that induce constitutive activity, regardless of ligand presence \citep{Spalding1998,Lu2000}. Maintaining the integrity of TM structure and sequence, then, is necessary for GPCRs to function properly.

As suggested by the strong biophysical, structural, and functional constraints imposed on GPCRs, one would expect that strong purifying selection dominates TM domain evolution. Alternatively, given the continued expansion of the GPCR gene family, notably of Rhodopsin family members such as olfactory receptors \citep{Lagerstrom2008,Niimura2003,Nei2007}, and of the array of ligands they receive, some positive selection should be detectable throughout GPCRs. As ligands tend to bind the TM, it is possible that positive selection there could drive the evolution of the GPCRs' expanding ligand repertoire. Here, we define positive selection as the ratio of the rate of nonsynonymous substitutions to synonymous substitutions, $dN/dS$, also known as $\omega$. When $\omega > 1$, positive selection may be inferred; alternatively when $\omega < 1$, there is evidence for purifying selection. Neutral evolution is indicated by $\omega = 1$.

Through a large-scale analysis of 359 mammalian GPCRs, we show that, on average, the TM evolves more slowly than does the EM, a result which should apply to all membrane proteins. Analysis of site rate variation across all GPCRs reveals that, unexpectedly, the average evolutionary rate heterogeneity of the TM is greater than that of the EM, in spite of the stronger biophysical and functional constraints the TM experiences. We additionally find evidence of positive selection in roughly half of the proteins studied here, in both their EM and TM domains. Chemosensory receptors, which includes all GPCRs (olfactory, taste, and vomeronasal receptors) that interact with exogenous chemical stimuli \citep{Mombaerts2004}, exhibit accelerated evolution relative to non-chemosensory GPCRs. This effect is highly pronounced in chemosensory GPCR TM spans. Finally, contradictory to previous reports \citep{Oberai2009}, we show that the lowered evolutionary rate of TM domains cannot solely be attributed to increased residue burial by other protein residues, but instead seems to stem from the membrane environment itself.

\section{Results}
\label{sec:2}
\subsection{Extracellular and intracellular domains evolve under similar selective pressures}
\label{sec:3}
We implemented the Goldman Yang codon evolutionary model (GY94) to estimate an average evolutionary rate $\bar{\omega}$ for each protein using the HyPhy batch language \citep{GoldmanYang1994,KosakovskyPondetal2005}. We compared fits between three models --- one with a single partition forcing both the TM and EM to evolve at an equal rate, one with two partitions (EM and TM), and one with three partitions (extracellular, TM, and intracellular). The latter two models allowed for each partition to have unique parameter values for $\bar{\omega}$, $\kappa$, $t$, and equilibrium codon frequency, where  $\kappa$ is the ratio of transition to transversion rates and $t$ is time, or branch length (see Methods for details). For each gene, the three models were compared using Akaike Information Criterion (AIC) scores such that models with lower AIC scores were preferred \citep{Akaike1974}. AIC scores are reported here as the difference of AIC scores ($\Delta \text{AIC}$) between two competing models, averaged across all genes. A larger $\Delta \text{AIC}$ indicates more support for the preferred model.

The two-partition model, on average across all genes, performed significantly better than the model which considered all domains as a single evolutionary unit ($\Delta \text{AIC} \sim 100$), and the three-partition model performed slightly better than the two-partition model ($\Delta \text{AIC} \sim 5$). However, there was no evidence that intracellular and extracellular regions had different average evolutionary rates in the three-partition model (paired t-test between extracellular and intracellular $\bar{\omega}$ values, $p = 0.589$). Therefore, the three-partition model was likely preferred due the marked difference in $\kappa$ between intracellular and extracellular regions (paired t-test between extracellular and intracellular $\kappa$ values, $p = 4.628\times10^{-07}$). Because no difference was detected between intracellular and extracellular $\bar{\omega}$, the two-partition model was used for all subsequent evolutionary rate analyses for all proteins. In terms of selection pressures, therefore, EM domains should be viewed as a single evolutionary unit. Our finding contradicts previous studies which claimed that intracellular regions of membrane proteins evolved more slowly than extracellular regions \citep{Julenius2006}. Our analysis shows no support for that hypothesis, likely due to our increased data sampling and more precise methodology; previous results may have been false positives.

\subsection{TM domains evolve more slowly than EM domains}
\label{sec:4}
\begin{figure*}[H]
\centerline{\includegraphics[width=4in]{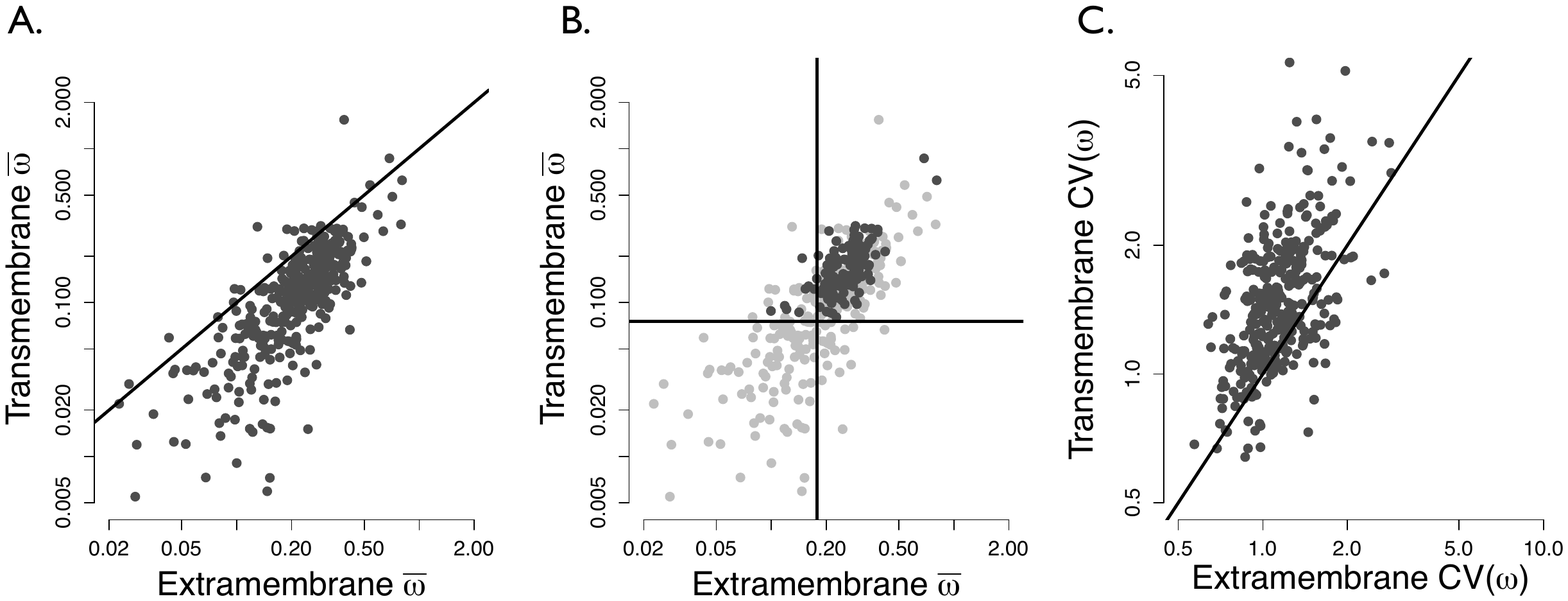}}
\caption{(\textbf{a}) TM $\bar{\omega}$ plotted against EM $\bar{\omega}$, both calculated by the REL model, shown on a log-log scale for all 359 proteins. The solid line indicates points where the EM rate equals that of the TM. The vast majority of proteins lie below this line (exact binomial test $p < 10^{-15}$), showing that TM domains evolve slower than EM domains. (\textbf{b}) Average TM $\omega$ plotted against average EM $\omega$, showing different types of GPCRs, on a log-log scale. Dark gray points represent chemosensory receptors, and light gray points non-chemosensory receptors. Vertical and horizontal lines lie at the mean of non-chemosensory TM and EM average rates, respectively. The majority of chemosensory points lie fall in the top-right quadrant of this plot (exact binomial test $p=6.85\times10^{-4}$), indicating their elevated evolutionary rate relative to non-chemosensory receptors. (\textbf{c}) Regression of TM against EM coefficients of variation of $\omega$ values [CV($\omega$)] on a log-log scale. If the spread of rates between partitions were equal, all points would lie roughly on the $x=y$ line shown. However, the majority of points lie on the TM side of the line, demonstrating the increased rate heterogeneity in TM domains of receptors proteins. This shift is highly significant at $p <10^{-15}$ by the exact binomial test.}
\label{fig:fig12}
\end{figure*}

We first broadly assessed rate differences between the evolution of TM and EM domains for each protein by estimating a single global $\bar{\omega}$ for each partition. Results from this analysis supported the hypothesis that, on average, EM regions evolve faster than their respective TM regions (Figure~\ref{fig:fig12}A). 94\% of the genes studied here (338 of 359) showed TM $\bar{\omega}$ values less than their gene's EM $\bar{\omega}$ (exact binomial test, $p < 10^{-15}$). A paired t-test comparing log-transformed EM and TM $\bar{\omega}$ values across each gene showed that EM rates are on average 0.094 greater than TM rates ($p < 10^{-15}$). We additionally found that the correlation between log-transformed EM rates and TM rates was highly significant ($r=0.75$, $p < 10^{-15}$), indicating that each protein likely has its own characteristic rate of evolution. 

\subsection{Elevated evolutionary rate in chemosensory receptors}
\label{sec:5}
Roughly one-third of receptors we analyzed were chemosensory receptors (127 of 359), of which 4 were taste receptors and the remainder olfactory receptors. We found that, relative to non-chemosensory GPCRs, chemosensory receptors exhibit significantly elevated evolutionary rates in both TM regions (t-test between log-transformed chemosensory TM and non-chemosensory TM $\bar{\omega}$ values, $p < 10^{-15}$) and EM regions (t-test between log-transformed chemosensory EM and non-chemosensory EM $\bar{\omega}$ values, $p < 10^{-11}$), as shown in Figure~\ref{fig:fig12}B. The $\bar{\omega}$ values for chemosensory receptor TM domains are, on average, $\sim 0.092$ greater than those of non-chemosensory receptors, and the $\bar{\omega}$ values for chemosensory receptor EM domains are, on average, $\sim 0.077$ greater than in non-chemosensory EM domains.  

To determine whether the TM or EM domains experience a greater evolutionary rate increase from chemosensory to non-chemosensory receptors, we compared the mean ratios of TM rate to EM rate between the two receptor types. We recovered a chemosensory ratio of 0.68 and a non-chemosensory ratio of 0.52 (independent samples t-test $p = 2.7\times10^{-8}$). That the chemosensory TM:EM rate ratio is, on average, significantly greater than the non-chemosensory TM:EM rate ratio demonstrates that the TM $\bar{\omega}$ increase from non-chemosensory to chemosensory receptors exceeds the EM $\bar{\omega}$ increase. Additionally, we performed a regression analysis with a TM $\bar{\omega}$ response and two predictors: EM $\bar{\omega}$ and receptor type (chemosensory or non-chemosensory). Both EM rates and receptor types have highly significant effects ($p < 10^{-15}$ and $p < 10^{-8}$, respectively) on TM rates. This result further supports our conclusion that TM rates increase more dramatically than do EM rates between non-chemosensory to chemosensory GPCRs.

We then examined whether it was more likely for TM or EM domains to exhibit a higher evolutionary rate in chemosensory receptors compared to non-chemosensory receptors. From an exact Fisher test, we recovered an odds ratio of 2.11 ($p=0.02$) in favor of the TM. This result demonstrates that 
chemosensory receptors are twice as likely to have elevated $\bar{\omega}$ in TM spans than in EM regions, compared to non-chemosensory receptors.

We further sought to examine whether the elevated evolutionary rate of chemosensory receptors could  be attributed to differential tissue expression. Indeed, evolutionary rates tend to be higher for proteins with a lower expression breadth, as may be the case for chemosensory receptors \citep{Duret2000,Liao2007,Pal2006}. Though it was once presumed that olfactory receptor expression was restricted to olfactory epithelium \citep{Buck1991}, recent studies have revealed that olfactory receptors are expressed in a multitude of diverse tissues in mammals \citep{Vanderhaeghen1997,Feldmesser2006,Zhang2007}. However, whether these receptors function in non-olfactory capacities is unknown. Thus, their activity may be limited to sensory tissue, which could cause their elevated evolutionary rates.

To assess the influence of expression breadth on evolutionary rate in GPCRs, we first obtained microarray expression data for 169 of our GPCRs from the Human Protein Atlas (http://www.proteinatlas.org) and regressed each gene's evolutionary rate on expression breadth and receptor type. We did not recover a significant relationship between evolutionary rate and expression breadth for G protein-coupled receptors (EM $p = 0.684$ and TM $p = 0.722$). However, the microarray data which we were able to collect was highly biased towards non-chemosensory receptors---only 12 of the genes for which we had expression data were chemosensory (1 taste and 11 olfactory). Therefore,  that limited amount of chemosensory expression data relative to non-chemosensory expression data may have biased our conclusions regarding the influence of expression breadth on $\bar{\omega}$. Possibly, then, chemosensory receptor expression breadth may contribute to their higher $\bar{\omega}$ values, but we lacked the statistical power to detect such an effect here.

\subsection{TM domains display increased rate heterogeneity}
\label{sec:6}
To assess evolutionary rate variation among sites, we calculated an $\omega$ for each residue of our 359 proteins using a random effects likelihood model (REL). From these rates, we determined the coefficient of variation for $\omega$ [CV($\omega$)] across partitions. We used CV($\omega$) as a proxy for rate heterogeneity. We found that the mean CV($\omega$) for TM domains was $0.402$ greater than for EM domains (paired t-test between each protein's TM and EM CV($\omega$) values, $p < 10^{-15}$). This increased spread of rates in the TM regions revealed their more extensive rate heterogeneity relative to their EM counterparts (Figure~\ref{fig:fig12}C). This effect holds for both chemosensory and non-chemosensory receptors.

\begin{figure*}[H]
\centerline{\includegraphics[width=4in]{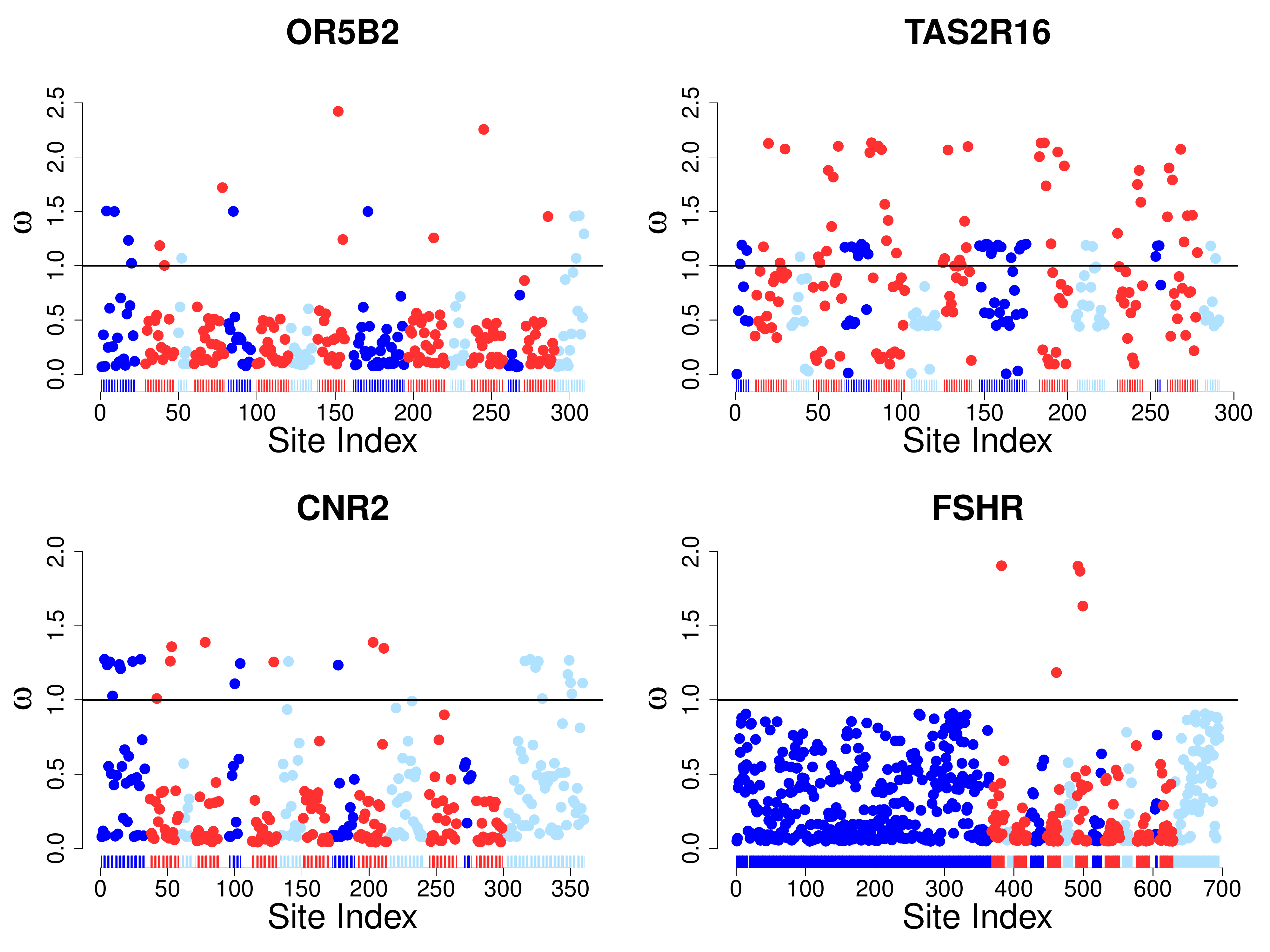}}
\caption{Distribution of residue $\omega$ values across four example proteins that contain positively selected residues. OR5B2 and TAS2R16 are both chemosensory receptors (olfactory and taste, respectively), CNR2 is cannabinoid receptor 2, and FSHR is the follicle-stimulating hormone receptor. Each of these proteins contain positively selected residues, in particular the chemosensory proteins. Red points represent TM residues, light blue points represent intracellular residues, and dark blue points represent extracellular residues, all as predicted by GPCRHMM \citep{Wistrand2006}. The bar at the bottom of the plots signifies the overall structure of the protein, with the same color-coding as the points. White spaces in between regions along the bottom bar indicate that the residue was not included in analysis, either due to lack of confidence in structure or alignment. Note that in each figure the entire structure of a GPCR is clearly visible - an extracellular N-terminus, intracellular C-terminus, three intracellular loops, three extracellular loops, and seven TM domains.}
\label{fig:site_plots}
\end{figure*}

While the majority of sites in GPCRs are under strong purifying selection, we identified 157 proteins (over two-fifths of our data set) which show evidence of positive selection at some sites. Positively selected sites were identified as those residues with an $\omega > 1$. Of all proteins analyzed, 31.5\% had EM residues with $\omega>1$, and 20.9\% had TM residues with $\omega>1$. Figure~\ref{fig:site_plots} depicts the selective regimes for several genes. 

To assess bias in the location of positively selected residues, we conducted a Cochran-Mantel-Haenszel Test, a stratified contingency table analysis of association, across all genes. Our overall contingency table was comprised of an array of $2\times2$ contingency tables for each gene, wherein each $2\times2$ table compared the number of positively and negatively selected sites in each partition. We recovered an overall odds ratio of 2.25 ($p < 10^{-15}$) in favor of EM. This result strongly suggested that positively selected residues were more than twice as likely to occur in the EM than in the TM. This trend held for both chemosensory and non-chemosensory receptors. Thus, even though there are more positively selected sites in EM domains relative to TM domains, we emphasize that positively selected residues are not uncommon in the TM. A list of all genes with positively selected residues can be found in accompanying Supplementary Information.

\subsection{Slowed TM evolution is not caused by structure}
\label{sec:7}
Finally, we assessed the extent to which structure influences the evolutionary rate in GPCR TM domains. For this analysis, we calculated each residue's relative solvent accessibility (RSA) from ten empirical crystal and one theoretical GPCR structure (see Methods for details). These structures represent all the currently known GPCR structures from the PDB. This effort was motivated by previous studies which have suggested that TM domains evolve slowly due to their relatively high percentage of buried residues \citep{Stevens2001,Oberai2009}. In this context, being buried refers to burial by other protein residues in the polypeptide, not by the plasma membrane itself. Buried residues are known to correlate strongly with a lower evolutionary rate \citep{FranzosaXia2009,Ramseyetal2011}. RSA directly measures how buried or exposed residues are within a protein structure, making it an ideal metric for this analysis.

\begin{figure*}[H]
\centerline{\includegraphics[width=4in]{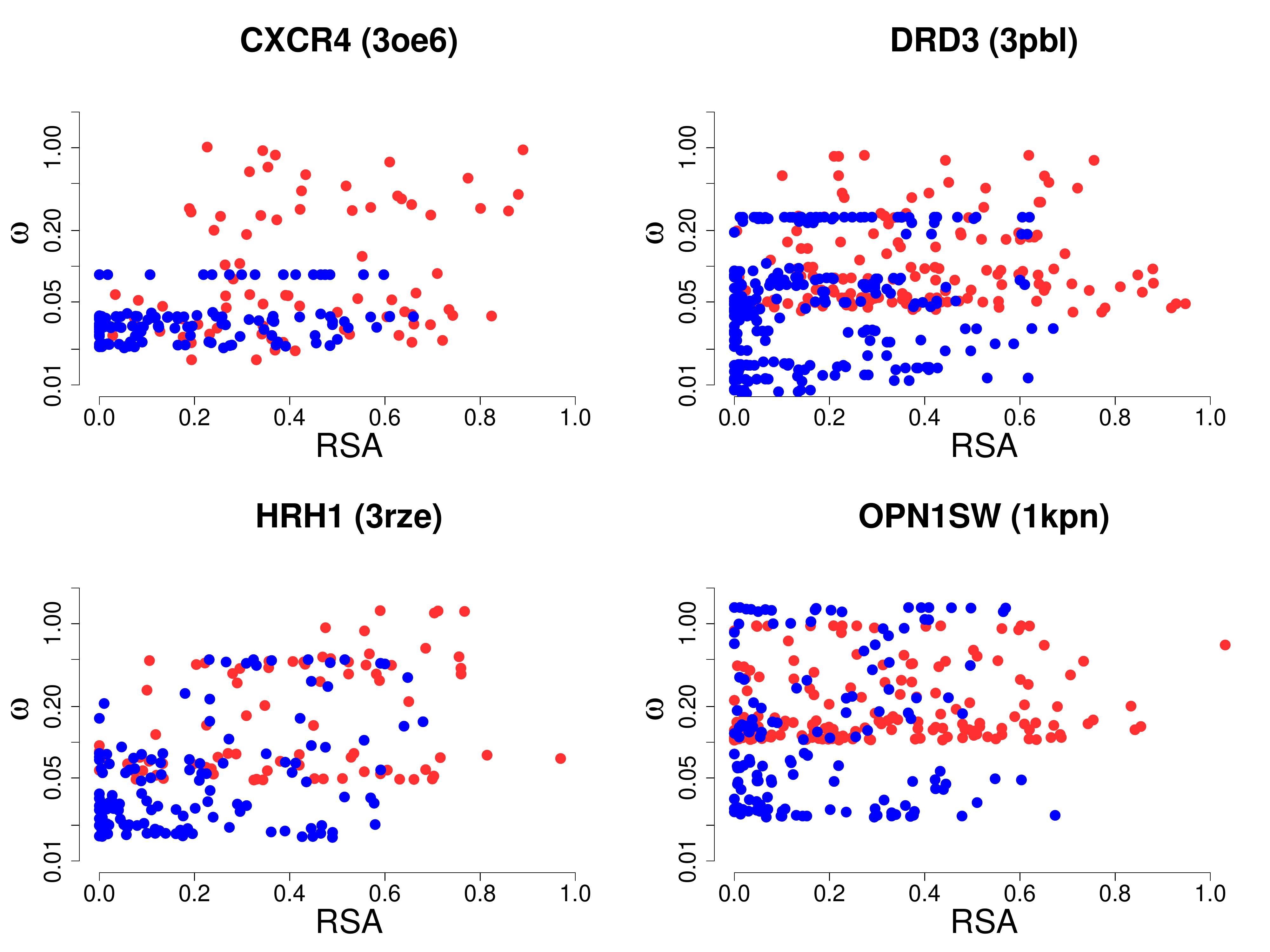}}
\caption{Regression of the $log(\omega)$ against RSA and residue partition. Red points represent TM residues and blue points represent EM residues. The gene name with its PDB ID in parentheses is shown above each graph. Linear regressions show that partition has a highly significant effect for each protein shown here (see Table 1 for more details). Further, the TM points display a noticeable shift towards higher RSA values, reflecting the increased burial by neighboring residues experienced in that domain.}
\label{fig:rsa_plots} 
\end{figure*}

After RSA was calculated for residues of the aforementioned eleven proteins, we regressed each residue's $\omega$ on RSA and partition (TM or EM). Results from this regression are shown in Table~\ref{tab:Tab1}. We systematically checked for interaction effects in each regression, and found that only two of the eleven proteins showed a significant RSA $\times$ partition interaction. Partition had a highly significant effect in eight of the remaining nine structures. These results demonstrate that the lowered rate of TM domains is not caused entirely by the higher percentage of buried residues they contain (Figure~\ref{fig:rsa_plots}), as had previously been hypothesized \citep{Stevens2001,Oberai2009}. Rather, it seems that the membrane environment, rather than protein structure itself, contributes to the lowered $\omega$ values characteristic of TM residues.

\section{Discussion}
\label{sec:8}

\begin{figure*}[H]
\centerline{\includegraphics[width=6in]{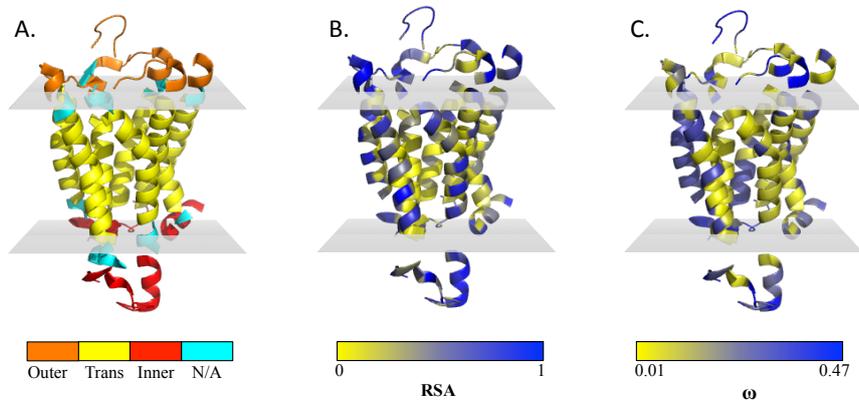}}
\caption{Structure of nociceptin receptor OPRL1 (PDB ID 4ea3), where gray planes represent borders of the plasma membrane. (\textbf{a}) Predicted extracellular, TM, and intracellular regions by GPCRHMM, which match the true structure nearly perfectly. Average TM $\omega$ for this protein is 0.0686, and average EM $\omega$ is 0.0837. Cyan residues marked `N/A' were excluded from analysis as GPCRHMM could not assign these residues with high confidence to a structural partition. These residues are not shown in parts B and C. (\textbf{b}) RSA (relative solvent accessibility) for each residue of OPRL1 analyzed. RSA values range from completed buried (0) to completed exposed (1). The vast majority of residues in the TM domain are buried whereas nearly all residues of the EM portions are highly exposed. (\textbf{c}) $\omega$ value at each residue of OPRL1 analyzed. Although each residue of this protein experiences purifying selection ($\omega<1$), rate heterogeneity is still pervasive throughout the protein.}
\label{fig:OPRL1}
\end{figure*}

We have demonstrated that the average evolutionary rate $\bar{\omega}$ of GPCR TM domains is significantly less than that of EM domains, mirroring results of previous studies which have suggested this trend across several types of membrane proteins \citep{Tourasse2000,Julenius2006}. Additionally, we have found that rate heterogeneity in TM spans exceeds that in EM regions and that many GPCRs experience positive selection across both structural domains. The average evolutionary rate of chemosensory receptors is also significantly greater than that of non-chemosensory receptors, specifically in the TM domains. Finally, we find no evidence, contrary to previous hypotheses, that increased residue burial influences the attenuated evolutionary rate of TM residues. Many of these results are summarized with a representative protein, the nociceptin receptor OPRL1, in Figure~\ref{fig:OPRL1}.

Although we found that the TM does evolve more slowly than does the EM, we emphasize that residues under positive selection were not uncommon across TM regions. Indeed, we identified 157 proteins, 55 of which are olfactory receptors, out of the 359 proteins we studied whose TMs contained residues with $\omega>1$. Thus, while biophysical constraints may have limited amino acid diversity in the TM, they did not preclude high rates of evolution at certain sites. Knowledge of positively selected sites within GPCRs may be useful for future biomedical research endeavors, as positive selection may be an indicator of a residue's functionality and potential use in drug development. A list of all GPCRs in this study with positive selected residues can be found in the Supplementary Information.

That TM rate heterogeneity exceeded EM rate heterogeneity was an unexpected result. Given the aforementioned structural and functional constraints, one might instead expect less variation across $\omega$ values of individual TM residues. Alternatively, while some key TM residues may experience  strong selective constraints, other residues will be much less important to protein structure and/or function. The former residues should be under exceedingly strong purifying selection, while the latter residues should be under weak purifying selection. In this dichotomy, there will be a strong difference in $\omega$ values between the highly constrained residues and the weakly constrained residues. In the EM, however, even the most constrained residues are, on average, under weaker negative selection  than are the most constrained TM residues. Thus, the difference between strongly and weakly negatively selected EM residues should be less than the difference between TM strongly and weakly negatively selected residues. Therefore, although somewhat unintuitive, the spread of evolutionary rates in the EM is smaller than in the TM.

Although other studies have previously investigated the evolutionary regimes in membrane proteins and olfactory receptors, our approach represents a dramatic methodological improvement. First, while previous studies of membrane proteins, including GPCR olfactory receptors, have focused either on ortholog duos or trios \citep{Tourasse2000,Clark2003,Julenius2006,Gimelbrant2004,Nielsen2005}, we have included up to 27 mammalian species per phylogenetic analysis (one phylogeny was created per gene). This increased breadth of species sampling should yield more robust conclusions. Specifically, we were able to infer the selective pressures at each residue rather than a single average $\omega$ for the whole protein. Had we not included that many species in our analyses, it would not have been possible to infer site-based evolutionary rates, the extent of rate heterogeneity, or positive selection at the residue level. Furthermore, previous studies of membrane proteins did not  conduct paired analyses, but rather compared average rates among all TM domains to average rates among all EM domains \citep{Tourasse2000,Julenius2006}. As we have demonstrated, there is a strong and highly significant correlation ($r=0.75$, $p < 10^{-15}$) between the TM and EM evolutionary rates within a single protein. Therefore, EM and TM $\bar{\omega}$ values within a single protein are not statistically independent, and a paired analysis as we have conducted is necessary to obtain statistically valid results.

Previous work has shown that transmembrane domains generally contain an increased proportion of buried residues relative to globular proteins or EM domains. This phenomenon is likely due to the highly packed arrangement of the TM span's constituent $\alpha$-helices \citep{Stevens2001,Oberai2009}. Typically, residue burial has been determined using the metric relative solvent accessibility (RSA), which measures the extent to which a residue in a protein structure is buried or exposed by other residues in the protein (not by the plasma membrane). Thus, RSA characterizes the local environment of a residue based on the extent of inter-residue contact, such that lower RSA values indicate increased burial by nearby protein residues. RSA is also a robust constraint on protein evolution, with buried residues evolving more slowly than exposed residues \citep{FranzosaXia2009,Ramseyetal2011}. It has thus been hypothesized that the  lowered evolutionary rate of TM domains could be attributed to their high percentage of buried residues \citep{Oberai2009}. 
Our evolutionary analysis of ten empirical and one theoretical GPCR structures, however, largely refutes this claim. We instead demonstrate that, while TM residues do display lower RSAs than do EM residues, this factor alone cannot explain the TM's lower evolutionary rate. Instead, we presume that the extreme biophysical constraints of the membrane environment as well as functional constraints are the leading factors which impose a lowered evolutionary rate on TM domains. As more empirical GPCR structures become available, this effect should be confirmed with larger data sets.

We have further demonstrated that chemosensory receptors exhibit increased rates of molecular evolution relative to other GPCRs. Although there are three main groups of chemosensory receptors (olfactory, taste, and vomeronasal receptors), we were only able to obtain mammalian orthologs for olfactory and taste receptors. As vomeronasal receptors specialize in detecting pheromones \citep{Mombaerts2004}, they should have highly species-specific sequences, thus making ortholog inference difficult. 

Previous studies on chemosensory receptor evolution have specifically investigated olfactory receptor evolution, the most common and diverse chemosensory receptors.  In general, olfactory receptors are one of most rapidly evolving gene families in human and other mammalian lineages \citep{Gilad2000,Clark2003,Nielsen2005}. Indeed, mammals contain at least 1000 olfactory receptors, and lineage-specific evolution of olfactory receptor families has been documented in primate splits \citep{Mombaerts2004,Gimelbrant2004,Gilad2005}. Although the olfactory receptor families are rapidly evolving, it has been suggested the the receptors themselves evolve primarily under weak purifying selection, and that there is no robust evidence for positive selection stronger than would be expected for any gene family \citep{Gimelbrant2004}. Our results indicate that, while weak purifying selection does dominate mammalian chemosensory receptor evolution, as noted by Gimelbrant et al. \citep{Gimelbrant2004} with regards to olfactory receptors, their average evolutionary rate is still significantly greater than the mean rate for their GPCR parent gene family. However, we also found that chemosensory receptors are not enriched for positively selected sites relative to other GPCRs, despite their increased $\bar{\omega}$. 

Given the rampant evolution of the number of olfactory receptors across species \citep{Niimura2003,Nei2007}, their elevated $\bar{\omega}$ was not unexpected. From an ecological standpoint, a mammal's ability to sense a diverse array of odorant and taste compounds is key for survival and species recognition. Such selection pressures are widely presumed to cause the high rate of olfactory gene turnover in animals, and we further this argument to include these genes' elevated rate of molecular evolution. The environmental selective pressures which cause frequent changes in the number of olfactory receptors likely also lead to the increased evolutionary rates of chemosensory receptors. Although both the TM and the EM domains evolve more quickly than do other GPCRs, we emphasize that the TM domains exhibit a more dramatic rate increase. This difference in protein domains could be explained by the ligand-binding pockets in chemosensory receptors. As both odorants and taste molecules bind chemosensory receptors within the TM region \citep{Mombaerts2004,May2006,Park2008,Lagerstrom2008}, positively selected residues in the TM span should broaden the diversity of odorants and tastes which mammals can sense. This widened diversity could contribute to key evolutionary processes, such as species recognition and speciation.

Based on our analysis of receptor protein evolution, we conclude that structural constraints do not always translate to constraints in evolutionary rate. Although biophysical considerations are important when assessing evolutionary parameters of different proteins, it should not be assumed that strong biophysical requirements limit a protein's ability to evolve, as reflected by the presence of positively selected residues in both the EM and TM. Our findings also shed light on the significant role that membranes play in constraining protein evolution, such that the hydrophobic environment imposes strong purifying selection on membrane proteins.

\section{Materials}
\subsection{Data collection and processing}
Human genes associated with the Gene Ontology annotation ``G protein-coupled receptor activity'' (accession GO:0004930) were collected from Ensembl Biomart. Using Ensembl's gene orthology prediction method \citep{Vilella2008}, we obtained orthologs from 27 other mammalian species with available genomes in the Ensembl database, and retained those sequences which contained no ambiguous residues. Subsequent analyses included all genes with at least 10 orthologs. Protein alignments were performed using Mafft within the Guidance package, to ensure high alignment quality \citep{Katoh2002,Penn2010}. As recommended by Privman et al. 2012 \citep{Privman2012}, we masked any residues in the resulting alignment with a guidance confidence score $< 0.9$ by changing their codons to ``NNN''. Phylogenies for each alignment were built using RAxML \citep{Stamatakis2006} with 100 tree inferences, and the resulting best tree was kept.

Each human protein sequence was partitioned into three structural partitions---intracellular, transmembrane, and extracellular domains---using the software package GPCRHMM, which gave individual posterior probabilities for each site belonging to one of those three partitions \citep{Wistrand2006}. Each site was categorized as either extracellular, intracellular, or transmembrane if its associated posterior probability was $\geq0.95$. All sites with posterior probabilities below $0.95$ were discarded. Each protein's partitions, as derived from the human sequence, were applied to all of its respective orthologs. Only genes with at least 50 amino acids per partition and whose TM comprised at least 15\% of their total length were kept. Additionally, any sequences with less than 40\% sequence identity to their orthologous human sequence were removed from alignments to ensure that all orthologs shared a common structure with the human protein. Positions corresponding to gaps in the human aligned sequence were removed. Sites belonging to each partition were concatenated such that each protein had a separate alignment for each region. Ultimately, 359 GPCR genes, averaging 18 sequences per alignment, were included in our analysis. Of these, 127 were chemosensory receptors (4 taste and 123 olfactory). 

\subsection{Evolutionary modeling to determine $\omega$ values}
We calculated the site-based evolutionary rate $\bar{\omega}$ for each protein with the HyPhy batch language, using the Goldman Yang codon evolutionary model (GY94) \citep{GoldmanYang1994,Yang2000,KosakovskyPondetal2005}. This Markov process model for codon substitution of $i$ to $j$ (for $i \neq j$) is given by the instantaneous rate matrix
\begin{equation}
Q_{ij} = \left\{ \begin{array}{rl}

              0                           &\mbox{more than one nucleotide changes} \\
              \pi_j                       &\mbox{synonymous transversion} \\
              \kappa\pi_j              &\mbox{synonymous transition}     \\
              \omega\pi_j              &\mbox{nonsynonymous transversion}  \\
              \kappa\omega\pi_j     &\mbox{nonsynonymous transition}
                            
                     \end{array} \right.,
\end{equation}
where $\pi_j$ is the frequency of codon $j$, $\kappa$ is the ratio of transition to tranversion substitutions, and $\omega$ is the ratio of nonsynonymous to synonymous substitution rates. The indices $i$ and $j$ include all 61 sense codons. The transition probability matrix additionally considered time, or branch length $t$, as measured by the expected number of substitutions for each codon across all residues \citep{GoldmanYang1994,Yang2000}. 

To begin, we calculated an average evolutionary rate $\bar{\omega}$ for each protein to infer the optimal partitioning strategy for analyzing TM v.s.\ EM evolution. In this case, the $\omega$ in our GY94 matrix corresponded to an average $\omega$ ($\bar{\omega}$) over all sites. Three models of protein evolution were examined; the first considered the entire protein a single evolutionary unit (single partition model), the second partitioned the protein into two distinct regions of transmembrane and extramembrane residues (two-partition model), and the third model partitioned the protein into three regions of transmembrane, intracellular, and extracellular regions (three-partition model). Models allowed each partition its own $\bar{\omega}$ , $\kappa$, and $t$ parameters. To identify the optimal number of partitions for GPCRs, we compared model fits with the Akaike Information Criterion \citep{Akaike1974}. AIC scores were calculated for each model of each gene and compared. The preferred model was the three-partition model. However, as there was no statistical difference between intracellular and extracellular $\bar{\omega}$  values in this model, the two-partition framework was used for all subsequent analyses.

We then implemented a random-effects likelihood (REL) model \citep{Yang2000, KosakovskyPondFrost2005}, again using the GY94 rate matrix, to discern an $\omega$ value for each residue across all proteins. In particular, we followed the RSA-independent model described in Meyer et al. \citep{Meyer2012}. To determine the optimal number of rate categories for each protein's partition, we ran the model 25 times, allowing the number of rate categories in each of the two partitions to vary from one to five in all possible combinations. AIC scores were calculated for each model, and the model with the lowest resulting AIC score was selected as the best-fitting model for that protein. 

To assign each site to a rate class, we employed an empirical Bayes approach \citep{NielsenYang1998} to calculate the posterior probability for each site belong to each rate class. Each site's rate was a weighted average over all rate classes by the associated posterior probability.
To calculate an average evolutionary rate $\bar{\omega}$ for each protein's partition, we took the weighted average, by the model's prior probabilities, of the $\omega$ values from each rate class. The standard deviation of $\omega$ values per partition was calculated using all residue $\omega$ values and the average $\omega$ value in a partition. Subsequently, we calculated the coefficient of variation for each protein's partition by dividing each partition's standard deviation of $\omega$ by its respective mean rate, $\bar{\omega}$.

\subsection{Structural analysis}
Relative solvent accessibility (RSA) was calculated for residues of 10 empirical and 1 theoretical GPCR structures obtained from the protein data bank (PBD). These PDB IDs, along with their respective gene names in parentheses, are 2rh1 (ADBR2); 3uon (CHRM2); 4daj (CHRM3); 3oe6 (CXCR4); 3pbl (DRD3); 3rze (HRH1); 4ej4 (OPRD1); 4ea3 (OPRL1); 1f88 (RHO); 3v2w (S1PR1); and theoretical structure 1kpn (OPN1SW). For each structure, we calculated the surface area for each residue using DSSP \citep{KabschSander1983} and normalized each value by its respective amino acid's maximum surface area value, as determined by Tien et.\ al.\ \citep{Tien2012}.

\begin{acknowledgements}
This work was supported by NIH grant R01 GM088344 to C.O.W. We thank Austin G. Meyer for his thoughtful comments.
\end{acknowledgements}

\bigskip

\begin{table}[H]
\caption { Results from the regression of log($\omega$) on RSA and partition (TM and EM) for each residue in 11 GPCR structures from the PDB. Significant values for partition and RSA are shown in bold. Empty values in the RSA $\times$ Partition area of the table indicate that no significant interaction effect was detected. }
\label{tab:Tab1}
\begin{tabular}{c c c c c c c c c}
\hline\noalign{\smallskip}
Gene Name & PDB ID & $r^2$ & Partition & p-value & RSA & p-value & RSA $\times$ Partition & p-value \\ 
\noalign{\smallskip}\hline\noalign{\smallskip}
ADBR2 & 2rh1 & $0.16$ & $\textbf{-0.28}$ & $0.0295$ & $\textbf{1.47}$ & $1.62 \times 10^{-6}$ &	& 	\\
CHRM2 & 3uon & $0.82$ & $\textbf{-1.58}$ & $<2 \times 10^{-16}$ & $0.12$ & $0.34$  &	& 	\\
CHRM3 & 4daj & $0.093$ & $\textbf{-0.19}$ & $3.61 \times 10^{-5}$ & $\textbf{0.68}$ & $1.02 \times 10^{-9}$ &	& 	\\
CXCR4 & 3oe6 & $0.26$ & $\textbf{-0.77}$ & $4.78 \times 10^{-8}$ & $\textbf{0.76}$ & $0.0069$ &	& 	\\
DRD3 & 3pbl & $0.14$ & $\textbf{-0.58}$ & $5.11 \times 10^{-8}$ & $\textbf{0.70}$ & $0.0024$ &	& 	\\
H1R1 & 3rze & $0.33$ & $\textbf{-0.81}$ & $1.99 \times 10^{-7}$ & $\textbf{1.77}$ & $8.51 \times 10^{-8}$ &	& 	\\
OPN1SW & 1kpn & $0.055$ & $\textbf{0.46}$ & $4.98 \times 10^{-4}$ & $0.49$ & $0.11$ &	& 	\\
OPRD1 & 4eje & $0.009$ & $\textbf{0.22}$ & $0.048$ & $0.16$ & $0.53$ &	& 	\\
OPRL1 & 4ea3 & $0.046$ & -$0.044$ & $0.647$ & $\textbf{0.91}$ & $5.91 \times 10^{-5}$ &	& 	\\
RHO & 1f88 & $0.017$ & -$0.084$ & $0.51$ & $0.30$ & $0.75$ & $\textbf{1.33}$ & $0.00467$\\
S1PR1 & 3v2w & $0.818$ & $2.93$ & $<2 \times 10^{-16}$ & $1.02$ & $1.66 \times 10^{-5}$ & -$\textbf{1.45}$ & $9.59 \times 10^{-5}$\\ 
\noalign{\smallskip}\hline

\end{tabular}

\end{table}


\end{document}